\documentclass[%
 amsmath,amssymb,
 reprint,%
]{revtex4-1}

\usepackage{graphicx}
\usepackage{dcolumn} 
\usepackage{bm} 

\usepackage[utf8]{inputenc}
\usepackage[T1]{fontenc}
\usepackage{mathptmx}
\usepackage{etoolbox}

\usepackage[table]{xcolor} 
\definecolor{myblue}{RGB}{25, 114, 177}
\definecolor{myorange}{RGB}{255, 126, 13}
\definecolor{mygreen}{RGB}{38, 157, 38}
\usepackage{ulem} 
\usepackage{multirow} 

\usepackage{lineno}
\definecolor{reviewer1}{RGB}{0,0,0} 
\definecolor{reviewer2}{RGB}{0,0,0} 
\definecolor{reviewerboth}{RGB}{0,0,0} 
\definecolor{reviewint}{RGB}{0,0,0} 


\begin{document}

\title[On the influence of reference sample properties on magnetic force microscopy calibrations]{On the influence of reference sample properties on magnetic force microscopy calibrations}

\author{Baha Sakar}
    \affiliation{Felix Bloch Institute for Solid State Physics, University of Leipzig, Linnéstraße 5, 04103 Leipzig}%
\author{Christopher Habenschaden}%
    \thanks{Corresponding Author}
    \affiliation{Physikalisch-Technische Bundesanstalt (PTB), 38116 Braunschweig, Germany}%
    \homepage{https://www.ptb.de/cms/en/ptb/fachabteilungen/abt2/fb-25/ag-252.html}
\author{Sibylle Sievers}%
    \affiliation{Physikalisch-Technische Bundesanstalt (PTB), 38116 Braunschweig, Germany}%
\author{Hans Werner Schumacher}
    \affiliation{Physikalisch-Technische Bundesanstalt (PTB), 38116 Braunschweig, Germany}%

\date{December 23, 2025}

\begin{abstract}
Magnetic force microscopy (MFM) allows the characterization of magnetic stray field distributions with high sensitivity and spatial resolution. Based on a suitable calibration procedure, MFM can also yield quantitative magnetic field values. This process typically involves measuring a reference sample to determine the distribution of the tip's stray field or stray field gradient at the sample surface. This distribution is called the tip transfer function (TTF) and is derived through regularized deconvolution in Fourier space. The properties of the reference sample and the noise characteristics of the detection system significantly influence the derived TTF, thereby limiting its validity range. In a recent study, the tip stray field distribution, and hence the TTF, of an MFM tip was independently measured in real space using a nitrogen vacancy center as a quantum sensor, revealing considerable discrepancies with the reference-sample-based TTF. Here, we analyze the influence of the feature distribution of the reference sample and the MFM measurement parameters on the resulting TTF. We explain the observed differences between quantum-calibrated stray field distributions and the classical approach by attributing them to a loss of information due to missing or suppressed spectral components. Furthermore, we emphasize the importance of the spectral coverage of the TTF. Our findings indicate that for high-quality reconstruction of the stray field of a sample under test (SUT), it is more critical to ensure a strong overlap of frequency components between the reference sample and the SUT than to achieve an accurate real-space reconstruction of the tip stray field distribution. 

---
Copyright 2025 Author(s). This article is distributed under a Creative Commons Attribution-NonCommercial-
NoDerivs 4.0 International (CC BY-NC-ND) License.

---
This is the author’s peer reviewed, accepted manuscript. However, the online version of record will be different from this version once it has been copyedited and typeset. PLEASE CITE THIS ARTICLE AS DOI: 10.1063/5.0288740

---
This article may be downloaded for personal use only. Any other use requires prior permission of the author and AIP Publishing. This article appeared in Review of Scientific Instruments 96, 123708 (2025) and may be found at \url{https://doi.org/10.1063/5.0288740}.
\end{abstract}

\maketitle

\section{\label{sec:INTRODUCTION}INTRODUCTION}


    Magnetic force microscopy (MFM) is a powerful technique for resolving nanoscale magnetic textures with high field sensitivity. \textcolor{reviewerboth}{With evolving instrumentation, developments in tip design (such as iron-filled carbon nanotubes \cite{Leonhardt2006, Wolny2008, Vock2010, Wolny2011}) and the implementation of advanced operation modes \cite{Schwenk2015, Zhao2018, Habenschaden2024, Sakar2021a, Neu2018}, field sensitivity and resolution nowadays reaches down to 80\,µT/$\sqrt{\textrm{Hz}}$ and 10\,nm \cite{Feng2022, Feng2022a, Necas2019, Corte‐Leon2020, Gisbert2021, Vock2011}.} 
    
    
    In common two-pass mode MFM, a magnetically coated tip mounted on an oscillating cantilever is scanned over the sample at a measurement height \( z \). The phase shift \( \Delta \phi \) of the oscillating cantilever is monitored to measure the interaction strength between the tip and the sample. For many applications, it is essential to obtain not only qualitative data but also quantitative data in terms of the sample's stray field distribution using either the magnetic field $H$ (in units of A/m) or the magnetic flux density $B = \mu_0 H$ (in units of Tesla). Such quantitative MFM (qMFM) requires calibrating the MFM system, that means finding the functional relationship that connects the magnetic field \( B^{\text{sample}}(x,y) \) of a thin film sample \textcolor{reviewer2}{(which can be modeled as magnetic surface charges projected onto the upper surface of the sample)} to the measured MFM signal, thereby providing spatially resolved phase shift data \( \Delta \phi(x, y) \).
    
    Initially, calibrations relied on interpreting \textcolor{reviewer2}{the magnetically coated etched silicon tips} as point-like dipoles or monopoles interacting with the local stray field, fitting this model to reference measurements. \textcolor{reviewer2}{While this approach is justified for nanotube-shaped tips, when working with coated tips,} these models exhibit significant limitations due to their strong dependence on the feature sizes of the sample \cite{Vock2010, McVitie2001, Schendel2000}. These limitations can be overcome by determining the spatial frequency-dependent transfer function of the MFM system, known as the instrument calibration function (ICF), as introduced by Hug et al. \cite{Hug1998} and Schendel et al. \cite{Schendel2000}. 
    
    \textcolor{reviewerboth}{To obtain the ICF, it is necessary to determine the magnetic stray field of the tip. In this paper, we demonstrate that while a high-quality reconstruction of the tip's stray field in real space appears desirable, it is more critical to ensure a strong overlap of frequency components between the reference sample and the SUT than to achieve an accurate real-space reconstruction of the stray field distribution.}
    
    \textcolor{reviewerboth}{The paper is structured as follows: In Section \ref{sec:mfm_cal-theory}, we introduce the basic theory of MFM calibration and compare a reference sample-based tip stray field calibration with an NV-based tip stray field calibration. Following this, we discuss the origins of the discrepancies between the real and reconstructed stray field distributions of the tip and their impact on qMFM measurements in Section~\ref{sec:METHODS}. To achieve this, we generate and analyze simulated MFM images of samples with well-defined features. This analysis will allow us, in Section~\ref{sec:refsamp_ttf}, to explore how the characteristics of the reference sample and measurement parameters influence the derived TTF. Additionally, we discuss the applicability of calibration based on different types of reference samples for calibrated measurements of various SUTs in Section~\ref{sec:ttf_strayfield}, showcasing the impact of (lacking) k-space overlap.}

\section{MFM Tip Calibration}\label{sec:mfm_cal-theory}

    \subsection*{Instrument Calibration Function and Tip Transfer Function}

\begin{figure}
        \centering
        \includegraphics[width=0.49\textwidth]{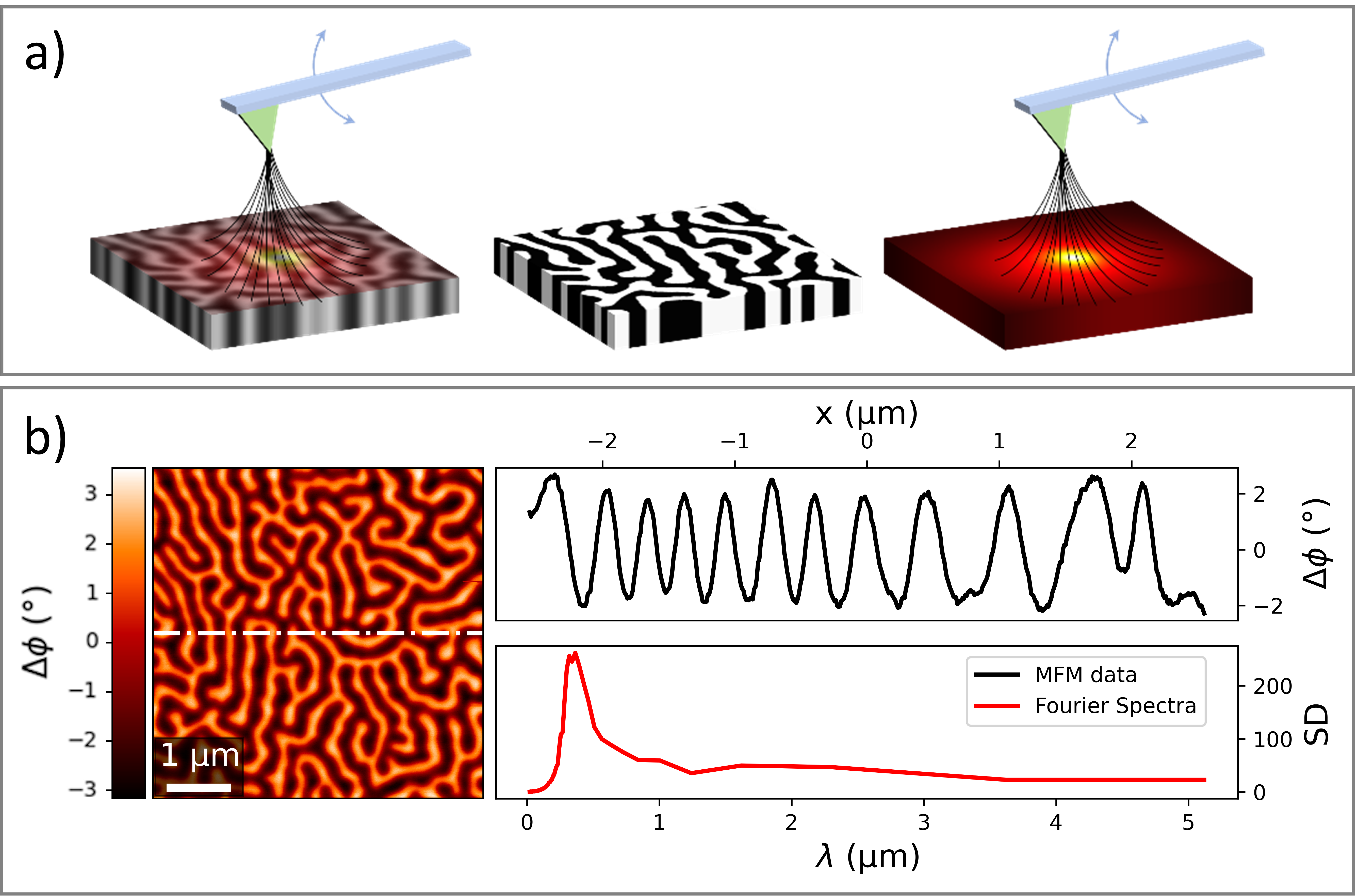}
        \caption{(a) MFM measurements represented as the convolution of the stray field from the tip with the sample magnetization. (b) MFM phase shift data for a 5.12\,µm x 5.12\,µm scan of a CoPt multilayer stack, displaying a maze domain pattern. A cross-section along white dashed is shown in black color. The corresponding wavelength density spectrum in Fourier space is shown in red color.}
        \label{fig:MFM_intro}
    \end{figure}

    The ICF is a function of the cantilever's mechanical properties and of the gradient of the force exerted on the tip. This force arises from the non-local interaction between the vertical component of the spatially extended MFM tip stray field distribution $B_z^{\text{tip}}$ and the magnetization of the sample under test (SUT), as illustrated in Fig.~\ref{fig:MFM_intro}a. The spatially extended tip stray field distribution at the sample surface thus governs the point-spread function character of the ICF.
    
    Mathematically, the measured cantilever phase shift can be conveniently described in partial Fourier space $(x, y, z) \to (k_x, k_y, z)=(\mathbf{k},z)$, where {$k = \sqrt{k_x^2 + k_y^2}$}. In this space, \textcolor{reviewer2}{the interaction between the tip's stray field and the sample's stray field can be represented by} the convolution of the tip's stray field at the sample surface with the \textcolor{reviewer2}{effective magnetic surface charge density} \cite{Corte‐Leon2020} of the sample, $\sigma_{\text{eff}}^{\text{sample}}$, which translates to a multiplication:

    \begin{eqnarray} \label{eq:phase_shift_intro}
        \Delta \phi(\mathbf{k}, z) =&& \text{ICF}(\mathbf{k},z) \cdot \sigma_{\text{eff}}^{\text{sample}} \nonumber \\
        \text{ICF}(k) =&& \frac{Q}{C \mu_0} \cdot \big( \text{LCF}(\mathbf{k}, \Theta, A) \big)^2 \cdot k \cdot B_z^{\text{tip}}(k, 0)
    \end{eqnarray}
    
    Here, $Q$ and $C$ represent the quality factor and the stiffness of the oscillating cantilever, respectively. The lever correction function (LCF) accounts for the finite oscillation amplitude $A$ and the canting angle $\Theta$ of the cantilever. Consequently, for a quantitative analysis of a SUT, it is essential to know both the shape and amplitude of the tip stray field distribution, as well as the scalar geometrical and mechanical parameters of the cantilever. This information allows for the calculation of the effective magnetic charge density distribution of any sample - and thus its stray field - by deconvolving the measured phase shift signal. Therefore, calibrating MFM requires determining $B_z^{\text{tip}}(k)$. By exploiting Eq.~\ref{eq:phase_shift_intro}, this can be achieved by measuring the phase shift distribution $\Delta \phi^{\text{ref}}$ of a reference sample with a well-known effective magnetic surface charge density $\sigma_{\text{eff}}^{\text{ref}}$. 
    
    Ideally, deconvolution to obtain $B_z^{\text{tip}}$ from Eq.~\ref{eq:phase_shift_intro} can be described as a division in Fourier space. However, due to measurement noise, this deconvolution is ill-posed and necessitates regularization, a method used to impose additional constraints to stabilize the inversion process and reduce noise in the reconstructed image. This is conveniently performed using an inverse Wiener filter (see Ref. \onlinecite{Wiener1949}) with a regularization parameter $\alpha_1$ (Eq.~\ref{eq:alpha_1}) as discussed in the relevant IEC standard on quantitative MFM \cite{IEC2021}. Typically, $\alpha_1$  is determined by an L-curve criterion \cite{Hu2020}. Due to  regularization, the reconstructed tip stray field distribution becomes frequency-filtered. To distinguish the reconstructed $B_z^{\text{tip}}$ from the actual $B_z^{\text{tip}}$, it is often referred to as the tip transfer function (TTF).

    \begin{equation} \label{eq:alpha_1}
        \text{TTF}(\mathbf{k}) = \frac{C}{k Q} \cdot \frac{\sigma_{\text{eff}}^{\text{ref}}(\mathbf{k})}{(\text{LCF}(k, \Theta, A))^2} \cdot \frac{\Delta \phi(\mathbf{k})^*}{|\Delta \phi(\mathbf{k}, z)|^2 + \alpha_1}
    \end{equation}
        
    Another factor influencing the TTF is the choice of the reference sample. To date, only \textcolor{reviewer2}{stripe} domain patterns have been tested and validated as suitable reference samples \cite{Hug1998,Hu2020,Sakar2021}. These samples are characterized by distinct domain patterns with specific domain sizes and narrow transitions between domains. 

    \textcolor{reviewerboth}{It is crucial to note that the calibration principle described here is based on the premise that neither the tip nor the reference sample is altered during measurements. Notably, the calibration sample must be measured in advance and subsequently re-measured after a SUT measurement to rule out any potential tip damage during the SUT measurement \cite{IEC2021}. A calibration is only applicable if these reference sample measurements coincide.}

    \subsection*{Tip Calibration using a Reference Sample}
    
    An MFM image of a typical \textcolor{reviewer2}{stripe} domain sample \textcolor{reviewer2}{(here a maze domain pattern that is composed of complex, interconnected pathways resembling a maze)}, specifically a CoPt multilayer stack as referenced in Ref. \onlinecite{Hu2020}, is illustrated in Fig.~\ref{fig:MFM_intro}b alongside a cross-section plot and its circularly averaged Fourier spectrum. 
    
    Determining the underlying magnetization pattern, which is necessary for calculating the effective surface charge density, involves discriminating the MFM image into up- and down-magnetized domains followed by applying a domain wall operator. The thereof calculated effective charge density pattern provides a good approximation of the real density pattern (see Fig.~\ref{fig:MFM_intro}) and is subsequently used in the deconvolution process.
    
    Although the fundamental validity of transfer function-based calibration has been demonstrated for these specific samples, its limitations - stemming from dependencies on the reference sample characteristics and measurement parameters such as image size and pixel resolution - have not yet been thoroughly investigated.

    Additionally, understanding the relationship between the TTF and the actual tip stray field distribution is crucial but challenging due to difficulties in directly assessing the tip stray field distribution with sufficient resolution and sensitivity. Over the past three decades, only a few attempts have been published that characterize the magnetic stray-field distribution of MFM tips using techniques such as the Hall effect \cite{Abelmann1998,Vock2014}, Lorentz tomography or holography \cite{Dai2018,Neu2018,Gao2004}. Recent advancements in NV microscopy (see Ref. \onlinecite{Casola2018}) have introduced new methods for highly spatially resolved characterization of magnetic stray fields.

    \subsection*{NV-measured Tip Stray Field Distribution}
    
\begin{figure}
        \centering
        \includegraphics[width=0.49\textwidth]{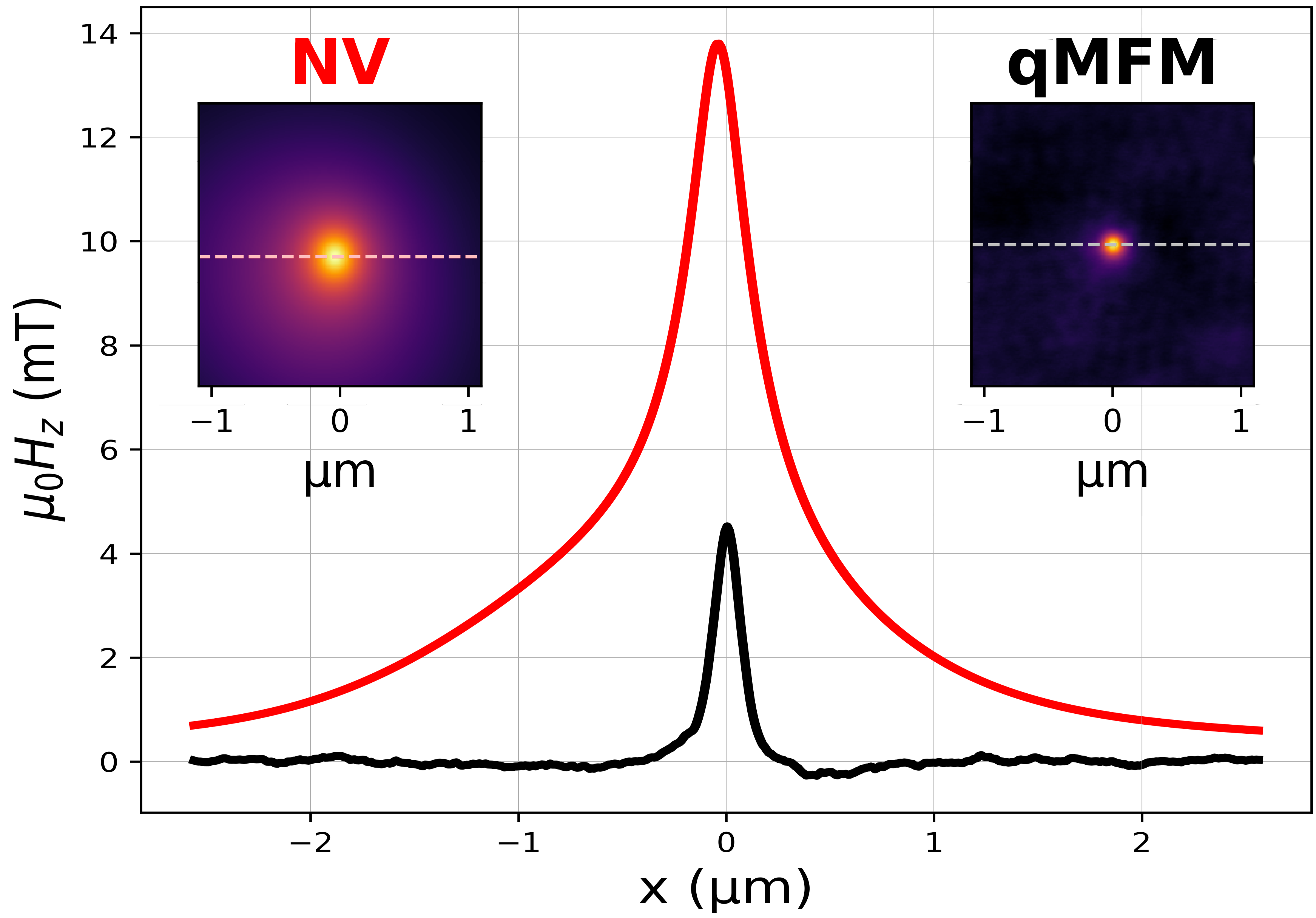}
        \caption{Comparison between \( \mu_0 H_z^{\text{tip}} \) measured by NV magnetometry (red) and qMFM calibration (black). The line profiles display a cross-section through \( \mu_0 H_z^{\text{tip}} \) as indicated by the dashed line in the insets.}
        \label{fig:ttf_nv_mfm}
    \end{figure}

    Exploiting these techniques, a recent study achieved a reference sample-independent determination of the tip stray field distribution by scanning an MFM tip over a single NV center embedded within a diamond solid immersion lens used as a quantum magnetic field sensor with quasi-atomic resolution \cite{Sakar2021a}. The MFM tip used in this investigation was a low-moment type (MFM\_LM, TipsNano), coated with 20\,nm CoCr and having a nominal tip radius of 30\,nm. The distance between the tip apex and the NV center was 80\,nm, while the measurement step size was set to 100\,nm. The collected data points were here fitted using a pseudo-pole tip model as described in Ref. \onlinecite{Haeberle2012}. Note, that in Ref. \onlinecite{Sakar2021a}, the fitting involved a Voigt function. The resulting NV measured \( B_z^{\text{NV}} \) distribution is illustrated by the red curve in Fig.~\ref{fig:ttf_nv_mfm}.
    
    The same MFM tip was subsequently characterized by a quantitative magnetic force microscopy (qMFM) calibration procedure at the same lift height of 80\,nm, employing a CoPt multilayer stack and the deconvolution process outlined in Eq.~\ref{eq:alpha_1}. This analysis yielded a reconstructed tip transfer function \( B_z^{\text{qMFM}} \), represented by the black curve in Fig.~\ref{fig:ttf_nv_mfm}.
    
    \textcolor{reviewerboth}{As introduced in Section \ref{sec:mfm_cal-theory}, it is essential to verify that no tip-sample interaction has altered the qMFM calibration. This verification was achieved by conducting repeated measurements with different tips in a NV $\rightarrow$ MFM $\rightarrow$ NV $\rightarrow$ MFM sequence, which yielded consistent results, indicating that no tip damage occurred. More information can be found in the supplementary materials of [\onlinecite{Sakar2021a}]. Fig.~\ref{fig:ttf_nv_mfm} presents one measurement from the verified series.}

    The first notable observation in Fig.~\ref{fig:ttf_nv_mfm} is the significant discrepancy between \( B_z^{\text{NV}} \) and \( B_z^{\text{qMFM}} \) regarding both amplitude and width of the stray field peak. This difference translates into a distinct divergence in their respective Fourier spectra. Specifically, the \( B_z^{\text{qMFM}} \) distribution is devoid of the low-frequency components that are predominant in determining the peak width.
    
    From the tip stray field distribution, the instrument calibration function (ICF) can be derived as discussed in Eq.~\ref{eq:phase_shift_intro}. Once the ICF is determined, the effective magnetic surface charge density \( \sigma_{\text{eff}}^{\text{sample}} \) of any SUT can be calculated from the MFM-measured phase shift data \( \Delta \phi^{\text{SUT}} \) through deconvolution. This process again requires regularization, executed as outlined in Eq.~\ref{eq:alpha_1}, using an inverse Wiener filter with a regularization parameter $ \alpha_2 $:
        
            \begin{equation} \label{eq:alpha_2}
                \sigma_{\text{eff}}^{\text{SUT}}(\mathbf{k}) = \frac{C}{k Q} \cdot \frac{\Delta \phi^{\text{SUT}}(k)}{(\text{LCF}(\mathbf{k}, \Theta, A))^2} \cdot \frac{B_z^{\text{tip}}(\mathbf{k}^*)}{|B_z^{\text{tip}}(\mathbf{k})|^2 + \alpha_2}
            \end{equation}
    
    The \( B_z^{\text{NV}} \) and \( B_z^{\text{qMFM}} \) data from the specific MFM tip used in Ref. \onlinecite{Sakar2021a} were utilized to calculate ICFs and subsequently reconstruct the stray field distribution of a CoPt multilayer reference sample using the MFM phase shift data measured with this MFM tip. Despite the pronounced differences between \( B_z^{\text{NV}} \) and \( B_z^{\text{qMFM}} \), the reconstructed stray field distributions showed good agreement, as illustrated in Fig. 4 of Ref. \onlinecite{Sakar2021a}. However, the reasons for this good agreement and its limitations have not been discussed in detail, yet.

\section{\label{sec:METHODS}Methods}
    
    To investigate the discrepancies between qMFM-derived and NV-derived stray field distributions, we generate artificial MFM images using forward simulations. These simulations utilize generic reference structures with well-defined characteristic features. In all cases, the image sizes $\Delta_{x,y}$, pixel sizes $\delta_{x,y}$, and resolution res$_{x,y}$ are equal in x- and y-directions. For easier reading, an image size of 5.12\,µm × 5.12\,µm will be referred to as an image size of 5.12\,µm. The same applies to the pixel size and resolution. We start from a typical measurement of a CoPt multilayer reference sample featuring a maze domain pattern. The pixel size was 10\,nm, and the image size was 5.12\,\textmu\text{m}. Furthermore, we assume that the \( B_z^{\text{NV}} \) data closely approximate the physical stray field distribution and thus can be considered as the "real" stray field distribution.
    
\begin{figure}
        \centering
        \includegraphics[width=0.49\textwidth]{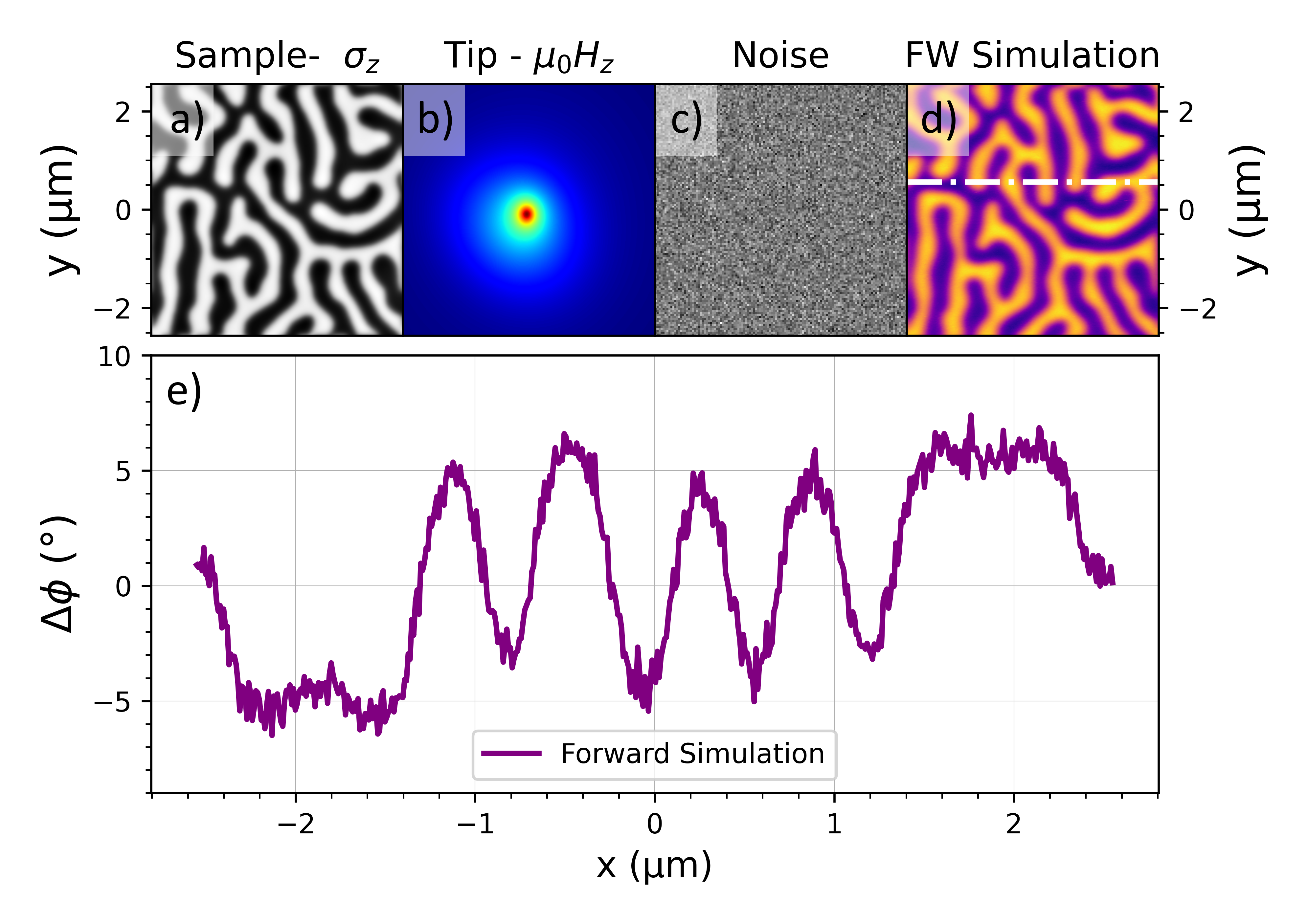}
        \caption{Description of forward MFM simulation resulting in (d) by convolving the field of a simulated sample (a) with the tip's field obtained from an NV measurement (b), and incorporating artificial noise (c). A line plot \textcolor{reviewer1}{(along the dashed line)} of the resulting simulated MFM signal is shown in (e).}
        \label{fig:forward_sim}
    \end{figure}

    Initially, we derive a binary magnetization pattern from the measured CoPt phase shift data (see Fig.~\ref{fig:MFM_intro}). Using this pattern, we calculate the effective charge density using well-established magnetic parameters of the CoPt stack. \textcolor{reviewer1}{The layer architecture of the CoPt stack is as follows: Pt(2 nm)/[(Co(0.4 nm)/Pt(0.9 nm)]$_{100}$/Pt(5 nm)/Ta (5 nm)/SiOx/Si(100), exhibiting a saturation magnetization of $M_s = (500 \pm 30)$\,kA/m, a Bloch-type domain transition with a domain wall width of approximately $\delta_{DW} =16$\,nm, and an average domain width of about 170\,nm \cite{Hu2020, Vock2014}}. \textcolor{reviewint}{Using these values, Fig.~\ref{fig:forward_sim}a can be derived.} To predict the expected MFM measurement response for the sample, we convolve the ICF, derived from \( B_z^{\text{NV}} \) data (Fig.~\ref{fig:forward_sim}b), with the calculated effective charge density pattern. Additionally, a typical noise pattern is incorporated into the simulation (Fig.~\ref{fig:forward_sim}c). This noise pattern is acquired from an ambient condition MFM signal measured over a non-magnetic sample using a commercial MFM. An example of such a derived artificial MFM phase shift distribution is depicted in Fig.~\ref{fig:forward_sim}d, with a cross-section illustrated in Fig.~\ref{fig:forward_sim}e. 
    

    To assess the influence of different magnetic feature sizes in reference samples and varying measurement conditions, we generate artificial reference samples, beginning with the original reference pattern image, which had 512 pixels, an image size of 5.12\,µm and consequently a pixel size of  10\,nm. We scale the original pattern to three larger sizes (10.24\,µm, 15.36\,µm and 20.48\,µm), creating three additional reference patterns with domain sizes larger than the original by factors of x2, x3, and x4, respectively. These are marked green in Table \ref{table1}. Due to this process, the pixel size increases to 20\,nm (x2), 30\,nm (x3) and 40\,nm (x4), respectively. To emulate measurements with varying resolution, we reduce the pixel size of all three previously discussed patterns to 10\,nm by a numerical interpolation. This process increases the number of pixels and thus changes the resolution (marked orange in Table \ref{table1}). Finally, we cut out a 5.12\,µm area from each of the latter high-resolution images, changing the image size while keeping domain size and pixel size (10\,nm) constant (marked blue in Table \ref{table1}). 
    
    We thus generated a set of artificial reference sample patterns with different domain sizes, resolution, and image sizes which will be used to analyze the impact of these parameters on the calibration process. 
    Using these datasets, we calculate emulated MFM data as described in the previous section following the approach depicted in Fig.~\ref{fig:forward_sim}a-d. Subsequently, the TTF is derived from each of the simulated noisy MFM images using the standardized approach. This involves discrimination, calculation of the effective charge pattern, and regularized deconvolution employing a Wiener filter (see Ref. \onlinecite{IEC2021}). In all cases, the regularization parameter for the Wiener filter is determined by applying the L-curve criterion to derive the optimal regularization and avoid ambiguities (for detailed methodology, see Ref. \onlinecite{IEC2021}). These simulated results are discussed in the following Section~\ref{sec:refsamp_ttf}.

\begin{table}
    \centering
    \caption{Properties of the artificially generated reference patterns. The colors correspond to the colors used in the plots in Fig.~\ref{fig:spectral_densities}. In all cases, the image sizes $\Delta_{x,y}$, pixel sizes $\delta_{x,y}$, and resolution res$_{x,y}$ are equal in x- and y-direction. }
    \label{table1}
    \setlength{\tabcolsep}{0pt} 
    \begin{ruledtabular}
        \begin{tabular}{ccccc} 
            Domain Size & Color & Image Size & Pixel Size & Resolution \\ 
              &  & $\Delta_{x,y}$ & $\delta_{x,y}$ & res$_{x,y}$ \\ 
            \hline
            \multirow{3}{*}{x2, s. Fig.~\ref{fig:spectral_densities}b} &  \textcolor{myblue}{\rule[0.5mm]{5mm}{2pt}}  & ~5.12\,µm & ~10\,nm & ~512\,px \\
             &  \textcolor{myorange}{\rule[0.5mm]{5mm}{2pt}}  & ~10.24\,µm & ~10\,nm & ~1024\,px \\
             &  \textcolor{mygreen}{\rule[0.5mm]{5mm}{2pt}}  & ~10.24\,µm & ~20\,nm & ~512\,px \\ 
            \hline
            \multirow{3}{*}{x3, s. Fig.~\ref{fig:spectral_densities}c} &  \textcolor{myblue}{\rule[0.5mm]{5mm}{2pt}}  & ~5.12\,µm & ~10\,nm & ~512\,px \\
             &  \textcolor{myorange}{\rule[0.5mm]{5mm}{2pt}}  & ~15.36\,µm & ~10\,nm & ~1536\,px \\
             &  \textcolor{mygreen}{\rule[0.5mm]{5mm}{2pt}}  & ~15.36\,µm & ~30\,nm & ~512\,px \\ 
            \hline
            \multirow{3}{*}{x4, s. Fig.~\ref{fig:spectral_densities}d} &  \textcolor{myblue}{\rule[0.5mm]{5mm}{2pt}}  & ~5.12\,µm & ~10\,nm & ~512\,px \\
             &  \textcolor{myorange}{\rule[0.5mm]{5mm}{2pt}}  & ~20.48\,µm & ~10\,nm & ~2048\,px \\
             &  \textcolor{mygreen}{\rule[0.5mm]{5mm}{2pt}}  & ~20.48\,µm & ~40\,nm & ~512\,px \\
        \end{tabular}
    \end{ruledtabular}
\end{table}

\section{\label{sec:refsamp_ttf}Influence of the Reference Sample on the Derived qMFM Tip Transfer Function}

\begin{figure*}
        \centering
        \includegraphics[width=0.99\textwidth]{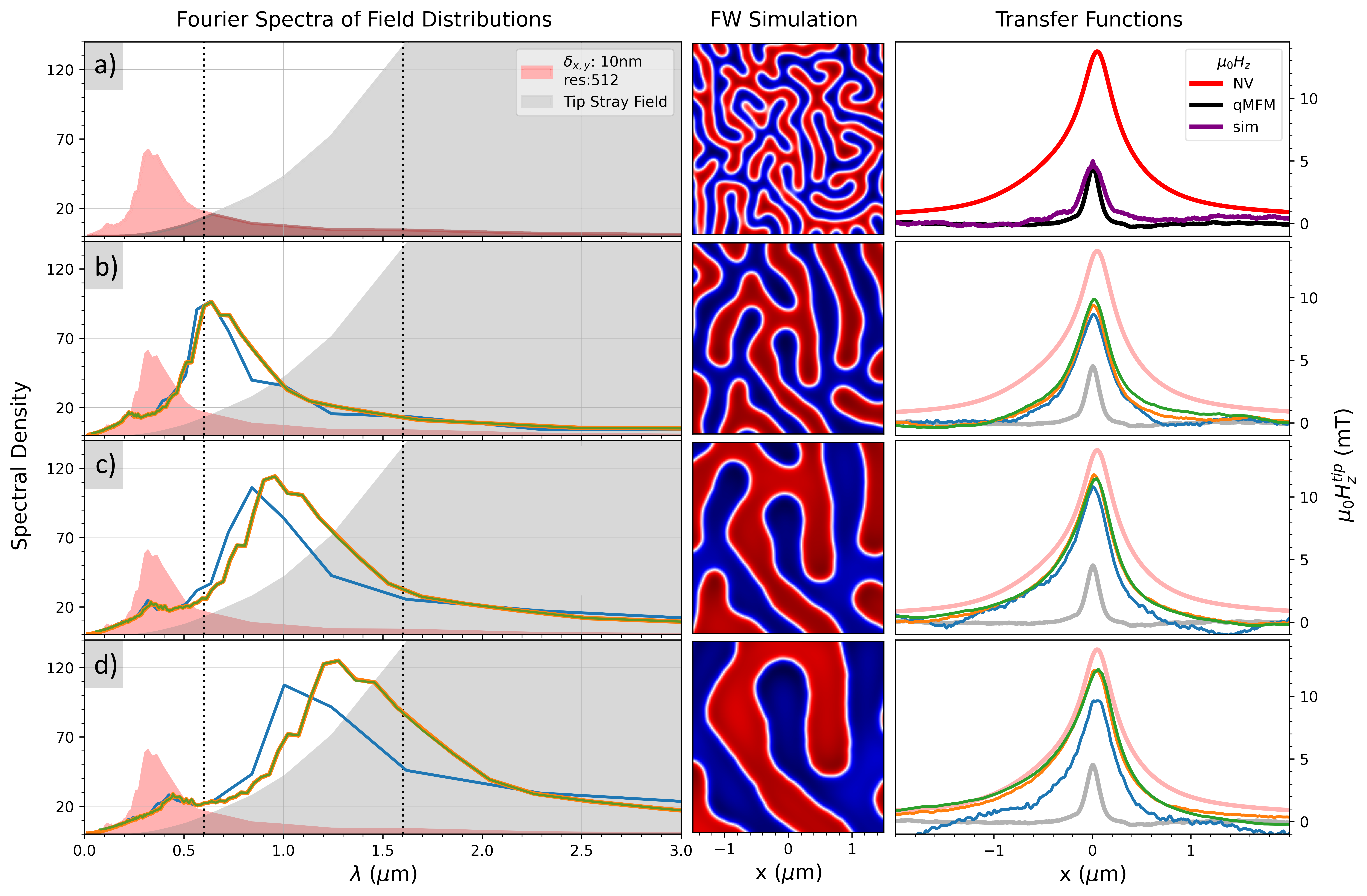}
        \caption{Comparison of spectral densities (left column) for different-sized reference patterns (middle column). The right column compares the $\mu_0 H_z^{\text{tip}}$ obtained from qMFM calibration using relevant reference patterns with the $\mu_0 H_z^{\text{tip}}$ measured by NV magnetometry. In (a), the spectral density of the CoPt sample's reference pattern, as measured with MFM, is compared to the spectral density of the magnetic tip measured by NV magnetometry. The $\mu_0 H_z^{\text{tip}}$ plotted in purple represents the field distribution obtained from a forward simulation using the original reference pattern as a control. From top to bottom, domain sizes of the reference patterns are increased by factors of 2 (b), 3 (c), and 4 (d). The blue, orange, and green plots correspond to different pixel resolutions and image sizes, see Table~\ref{table1}.}
        \label{fig:spectral_densities}
    \end{figure*}

    The results of simulated TTF analyses for various measurement and reference sample parameters are summarized in Fig.~\ref{fig:spectral_densities}. The TTFs presented in real space show circularly averaged data.
    
    The left panel of Fig.~\ref{fig:spectral_densities} shows the Fourier spectral density of the artificially generated reference patterns with different parameters, along with the Fourier spectrum of $ B_z^{\text{NV}} $ (grey) and the Fourier spectrum of the CoPt multilayer reference sample (light red). Note that the spectra are plotted against wavelengths $ \lambda $ in microns (feature size) instead of wave vector. \textcolor{reviewint}{The middle panels display the simulated reference patterns at a fixed pixel size of 10\,nm and a resolution of 512\,px.}
    
    The right panel shows the real-space representation of $B_z^{\text{NV}}$ (colored red) and the experimentally derived $B_z^{\text{qMFM}}$ (colored black), along with the real-space representations of TTFs extracted from simulated reference measurements using various reference patterns. Additionally, the right panel of Fig.~\ref{fig:spectral_densities}a shows the TTF calculated from a simulated measurement of the original reference sample as described (purple). This initial simulation serves as the control. The good agreement between measured and simulated TTFs confirms that the underlying model describes the MFM calibration process reasonably well, indicating the validity of our forward simulations for differently-sized reference patterns. 
    
    The spectral density plots of the original reference pattern and the tip field distribution $B_z^{\text{NV}}$ \textcolor{reviewint}{in Fig.~\ref{fig:spectral_densities}a} exhibit only a small overlapping area. This overlap (or the lack thereof) should limit the spectral information that can be reconstructed in any qMFM calibration. Specifically, a narrow Fourier spectrum of the reference sample should limit the Fourier components of the tip stray field distributions that are accessible through the calibration. Such limitations constrain both the validity, range, and quality of the reconstructed tip stray field distribution.
    
    To validate this hypothesis, we compare results from a standard CoPt reference sample (Fig.~\ref{fig:spectral_densities}a) with those obtained from TTFs derived via simulated calibrations using the artificially generated reference samples as described in Section \ref{sec:METHODS}. These simulated scans (see Table~\ref{table1}) exhibit larger domains, varied scan image sizes, and different resolutions, thereby altering their Fourier spectra compared to the original CoPt patterns. 
    
    In Fig.~\ref{fig:spectral_densities}b to d, \textcolor{reviewint}{the colors used in the figure correspond to those listed in Table~\ref{table1}.} Consequently, the blue curve represents data with a smaller image size, while the orange and green curves correspond to larger image sizes. Among these, the orange plot displays higher resolution (i.e., smaller pixel size) compared to the green curve. Generally, the average feature size of the reference patterns increases from Fig.~\ref{fig:spectral_densities}b to d. Note that the orange and green curves overlap almost exactly in all right-hand side panels. They differ only in the pixel size (i.e. image resolution) of the simulated measurement of the reference sample, suggesting that within the considered frequency range, resolution is of minor relevance for reconstructing the stray field distribution.
    
    As seen in the insets in the middle column of Fig.~\ref{fig:spectral_densities}, the feature size of the reference sample increases from a to d. Consequently, the spectrum's maximum for the orange/green curves and of the blue curve shifts to the right and thus a larger value of $\lambda$. Simultaneously, as $\lambda$ increases the TTF depicted on the right panels increasingly resembles the NV-measured field distribution. This trend supports our assumption: as the overlapping area between the reference sample's Fourier spectrum and the tip stray field distribution expands, the reconstructed TTF more closely aligns with the NV-measured tip field (colored red).
    
    Note, that for the largest feature size shown in Fig.~\ref{fig:spectral_densities}d, there is a significant difference between the TTF of the green/orange and the blue curve. This results from the up to four times smaller scan size of the blue data. Reducing the image size (blue curve) while maintaining domain size leads to a loss of large wavelength features, impeding an accurate reconstruction of the tip stray field. Additionally, for larger domain patterns, smaller images are less representative as they encompass only a few domains, capturing less of the stochastic variations inherent in the pattern.
    
    To better understand this behavior, we need to discuss the influence of noise. For ideal images without noise, it would be possible to recover the FFT even for spectral components with very low amplitudes (unless the components are zero). Since we added noise to the simulated MFM image, to better reflect real measurements, the deconvolution that gives the TTF must be regularized to suppress noise amplification. This leads to a suppression of frequency components with low amplitudes in the MFM image. Accordingly, we can sharpen the overlapping area criterion: The product of the ICF and the sample Fourier spectrum must be significantly higher than the noise level for a specific k-value. Otherwise, it will be suppressed by the inverse Wiener filter.
   
    Following this analysis, experimentally derived TTF data from two distinct reference samples are presented in the next section to validate our findings.

    \subsection*{\label{sec:experimental_validation}Experimental Validation}
    
\begin{figure}
            \centering
            \includegraphics[width=0.49\textwidth]{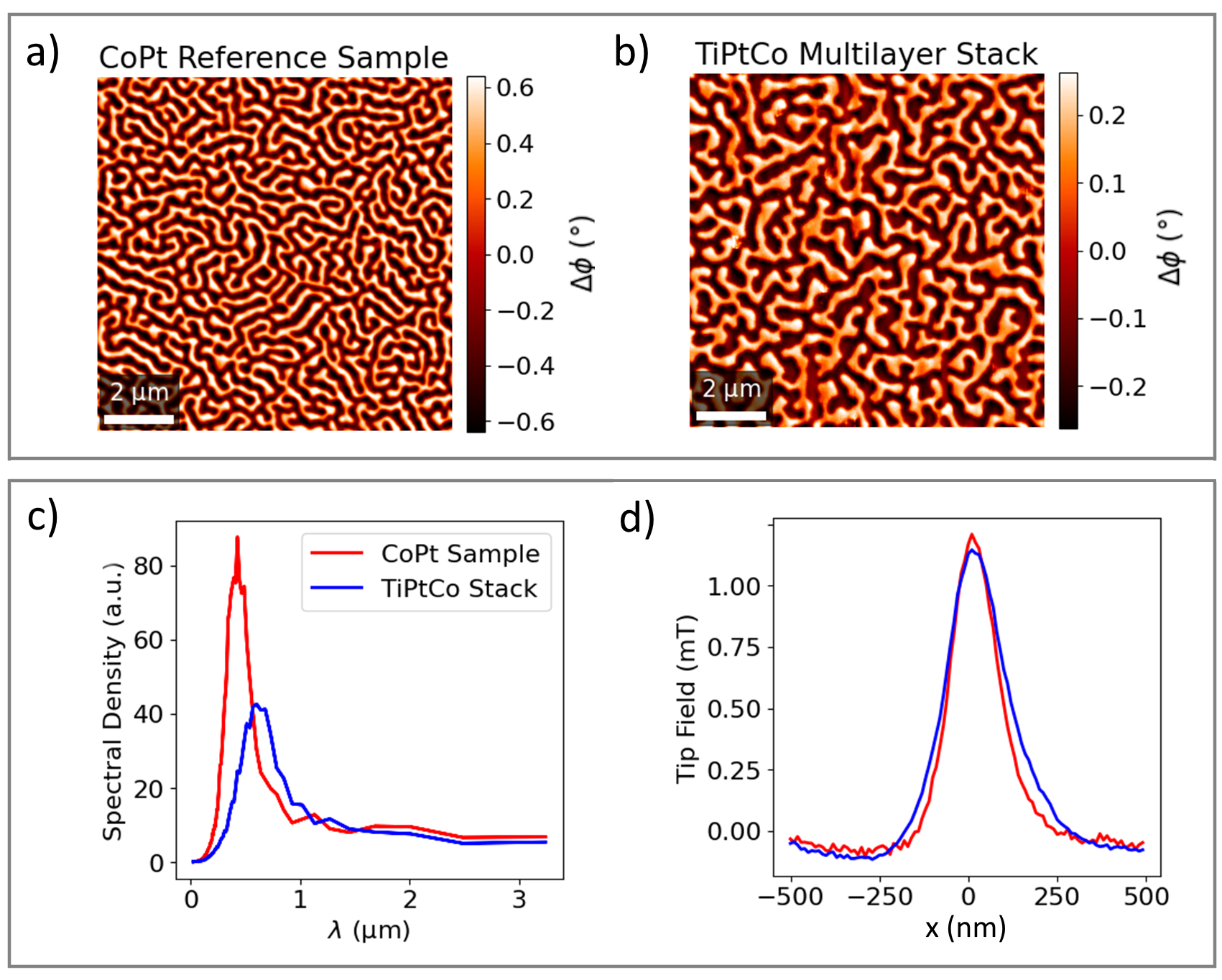}
            \caption{Comparison of TTFs derived from calibrations based on MFM measurements of two different reference samples. The CoPt reference sample (a) and the TiPtCo reference sample (b) exhibit distinct domain pattern widths, which are also reflected in their circularly averaged Fourier spectra (c). The TTF \textcolor{reviewer1}{(here visualized by a cross-section)} derived from the TiPtCo sample is notably wider than that obtained from the CoPt sample (d).}
            \label{fig:comparison_ttf}
        \end{figure}

        To determine if our findings are reproducible under experimental conditions, we compare TTFs derived from calibration measurements of two distinct reference samples. These samples include the previously discussed CoPt reference sample and a TiPtCo multilayer stack detailed in Ref. \onlinecite{Sakar2021}. Both exhibit maze domain patterns but with differing characteristic domain widths, as illustrated in Fig.~\ref{fig:comparison_ttf}a and b.
        
        \textcolor{reviewint}{The domain width of the TiPtCo stack is $D_{\text{TiPtCo}} = 345\,\text{nm}$, whereas the domain width of the CoPt stack is $D_{\text{CoPt}} = 235\,\text{nm}$. These widths were calculated using a self-correlation approach (see Ref. \onlinecite{Sakar2021}).}  Consequently, this results in different Fourier spectra, depicted in Fig.~\ref{fig:comparison_ttf}c, where the peaks of the TiPtCo magnetic surface charge distribution are shifted to higher wavelengths $\lambda$. \textcolor{reviewer1}{This Fourier spectrum was calculated by Fourier-transforming the two-dimensional measurements shown in Fig.~\ref{fig:comparison_ttf}a and b, followed by circular averaging of the resulting Fourier images. Thus, all Fourier components in all directions are considered.}

        The TTFs were calculated from these measurements as previously discussed. Given the different Fourier spectra of the samples, it was anticipated that using the same tip field would yield a somewhat wider TTF in real space after calibration with the TiPtCo multilayer stack. This expectation is confirmed by the \textcolor{reviewer1}{cross sections (which are taken in the x-direction through the center of the resulting TTFs, analogous to those in Fig.\ref{fig:ttf_nv_mfm})} shown in Fig.\ref{fig:comparison_ttf}d. The comparison reveals a wider TTF for calibrations using the TiPtCo sample, which has a broader domain width. This effect was not evident in Ref. \onlinecite{Sakar2021}, where stray field gradients were analyzed, which are less sensitive to higher \( k \)-components due to their slower decay behavior and consequently lower gradients. The change in the TTF is less pronounced than the differences observed in Fig.~\ref{fig:comparison_ttf}. We primarily attribute this to a lower ratio of domain widths, of only $D_{\text{TiPtCo}}/D_{\text{CoPt}} = 1.47$, and the use of a different tip, here. Furthermore, the lower stray field amplitude of the TiPtCo sample, combined with a low moment tip, results in a reduced signal-to-noise (S/N) ratio. Despite of these differences, we regard the results as at least a qualitative confirmation of our simulation-based findings.


        It is important to note that these samples not only differ in domain width but also in thickness and saturation magnetization, leading to a lower stray field amplitude for the TiPtCo sample. \textcolor{reviewer1}{In our experience, the CoPt sample can withstand medium moment tips without exhibiting tip-sample interaction; however, this is not the case for the TiPtCo sample, which leads to domain movement and an uneven distribution of dark and bright domains. To avoid modifications to the TiPtCo sample during imaging, a Nanosensors SSS-QMFMR low-moment tip was used for this measurement. Both samples were subsequently measured using the same tip in a Park Systems NX-Hivac MFM under ambient conditions at a lift height of 60\,nm. This class of low-moment tips has previously been verified to not cause tip-sample interaction with the CoPt sample \cite{Hu2020}. The TiPtCo sample has been demonstrated to withstand a MFM\_LM tip \cite{Sakar2021}, which has a higher moment than the SSS-QMFMR tip, making the latter a safe choice for this measurement.}

\section{\label{sec:ttf_strayfield}Influence of the Derived qMFM Tip Transfer Function on the Reconstructed Stray Field Distribution in qMFM}

    So far, we have analyzed how different Fourier spectra in MFM calibration patterns affect the resulting TTFs. In this section, we explore the impact of these derived TTFs on qMFM measurements of stray field distributions for typical magnetic nanostructures. The ICFs are calculated for the TTFs obtained from the simulated reference patterns. By deconvolving the simulated MFM test data with these ICFs, we derive surface charge density patterns and field distributions of the test simulations. These results are then compared with the fields calculated from the sample magnetization used in the simulation. 

\begin{figure}
        \centering
        \includegraphics[width=0.49\textwidth]{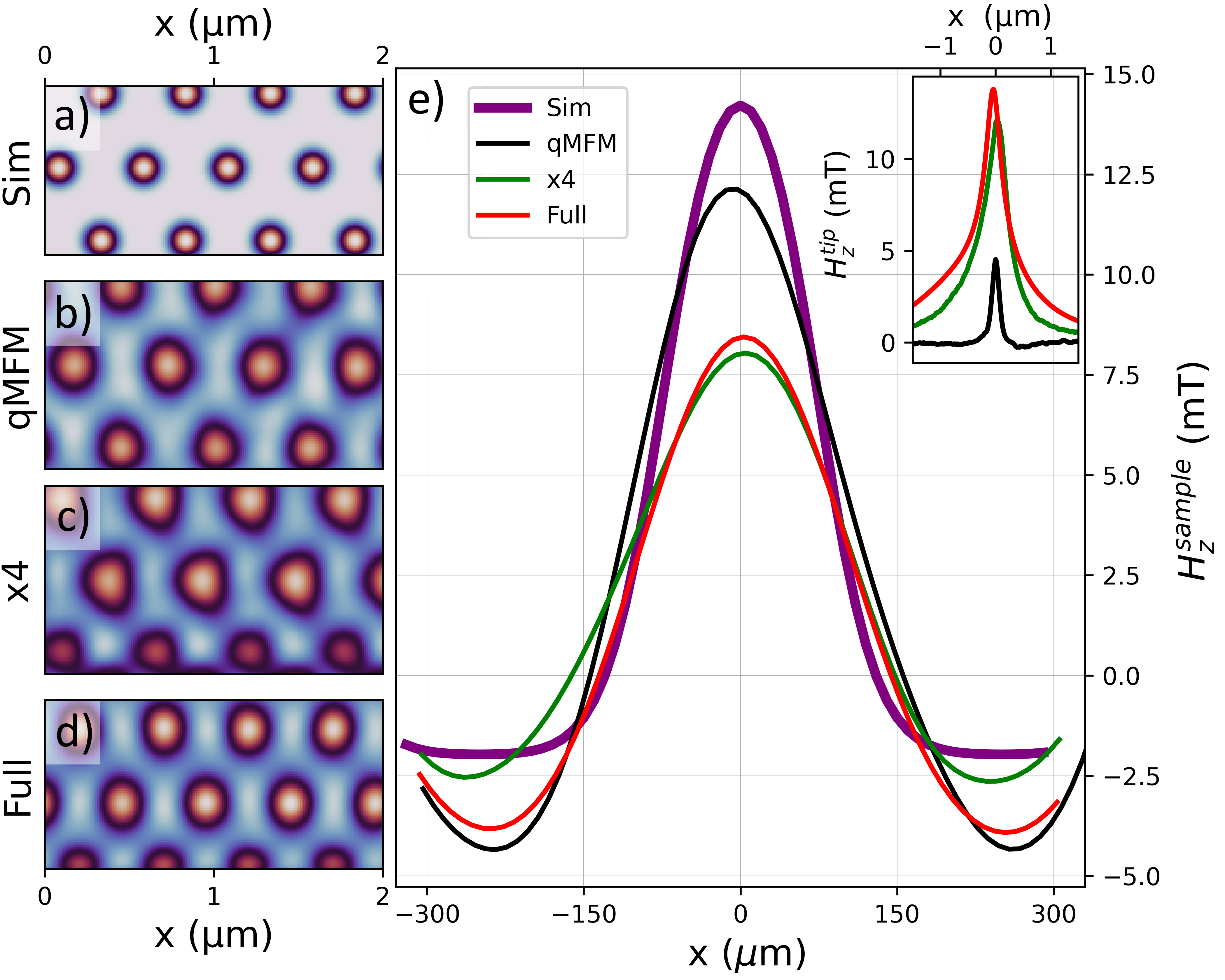}
        \caption{Artificial skyrmion sample. The left column shows the two-dimensional $\mu_0 H_z$ distributions of skyrmions obtained by simulation (a), deconvolution using the transfer function from the original reference pattern (b), and from the largest reference pattern (x4, c). The field in (d) is deconvolved using the distribution obtained from NV measurements. In (e), the plot lines representing the maximum values of the skyrmion field distributions are compared. The inset figure provides the transfer functions used for these deconvolutions.}
        \label{fig:validation_skyrmion_lattice}
    \end{figure}

    
    The two ``typical'' test samples represent structures with different characteristic length scales and are shown in Fig.~\ref{fig:validation_skyrmion_lattice}a and \ref{fig:validation_qr_structure}a, respectively. Fig.~\ref{fig:validation_skyrmion_lattice}a shows an artificial skyrmion lattice with a skyrmion diameter of 100\,\text{nm}, whereas  Fig.~\ref{fig:validation_qr_structure}a shows a micrometer-scale QR code-like structure. The skyrmion lattice was generated using the Romming formula \cite{Romming2015}. For the QR code, a purely perpendicular magnetization configuration with alternating up and down magnetized areas (light/dark shades) was assumed. For both simulated samples, parameters included a film thickness of $t = 25\,\text{nm}$ and a saturation magnetization of $M_\text{S} = 1.4 \times 10^5\,\text{A/m}$. From these magnetization distributions, the effective magnetic surface charge density and the sample's stray field at any distance were calculated.
    
    The simulated MFM phase shift measurement data were generated by convolving the effective magnetic surface charge density with the ICF calculated from the NV-measured $B_z^{\text{NV}}$ distribution, which represents the ``real'' tip stray field distribution. A cantilever stiffness of $C = 3.063\,\text{N/m}$, and a Q-factor of the cantilever of $Q = 160.7$ were used. The stray field was calculated at a distance of 80\,\text{nm} from the sample surface.
  
    

    The simulated MFM phase shift data were then deconvolved using different ICFs derived from the TTF data obtained for various artificial reference samples, as discussed in Section \ref{sec:refsamp_ttf}. We compared these fields with those calculated directly from the magnetization patterns used to simulate the test samples.

    Additionally, we also deconvolved the MFM data by:
    (i) Using the TTF and corresponding ICF obtained from experimental MFM measurements (labeled as ``qMFM'' in Fig.~\ref{fig:validation_skyrmion_lattice} and Fig.~\ref{fig:validation_qr_structure}), and
    (ii) Utilizing the ICF derived from $B_z^{\text{NV}}$ data (labeled ``Full'' in Fig.~\ref{fig:validation_skyrmion_lattice} and Fig.~\ref{fig:validation_qr_structure}). The field calculated directly from the magnetization pattern is denoted as ``Sim''.
    This resulted in eleven sets of stray field distributions. In Fig.~\ref{fig:validation_skyrmion_lattice} and Fig.~\ref{fig:validation_qr_structure}, we summarize results for selected TTFs derived from large reference patterns with the highest resolution to focus on characteristic behavior (marked orange in Table \ref{table1}). The curves of all other results, which confirm these findings but were omitted for better plot visibility, are not shown. 
    
    \subsection*{Artificial Skyrmion Sample}
    
    Fig.~\ref{fig:validation_skyrmion_lattice} illustrates the derived stray field data $\mu_0 H_z$ for the skyrmion lattice. The simulated stray field distribution exhibits the highest field amplitude, whereas all deconvolved data show reduced amplitudes, indicating a loss of information.

    The most accurate result is obtained from deconvolution using the experimental TTF (Fig.~\ref{fig:validation_skyrmion_lattice}b and Fig.~\ref{fig:validation_skyrmion_lattice}e – black). Conversely, the stray field distribution derived from $\mu_0 H_z^{\text{tip}}$ using a reference pattern four times larger is the least accurate (Fig.~\ref{fig:validation_skyrmion_lattice}c and Fig.~\ref{fig:validation_skyrmion_lattice}e – green). This discrepancy arises due to the absence of higher frequency $k$-vectors in the Fourier spectrum of the reference pattern.
    
    The field distribution deconvolved using $\mu_0 H_z^{\text{tip}}$ measured by NV is similarly poor (Fig.~\ref{fig:validation_skyrmion_lattice}d and Fig.~\ref{fig:validation_skyrmion_lattice} – red). This suggests that the interpolation of $\mu_0 H_z^{\text{tip}}$, achieved by fitting NV data with a multipole model, failed to accurately reconstruct higher frequency $k$-vectors in the Fourier spectrum. These vectors are predominantly induced by magnetization in the tip-apex area. We assume, that the 100\,nm grid used in the NV measurement leads to a loss of information. Furthermore, the significant distance between the tip and NV during the measurement of $B_z^{\text{NV}}$ likely resulted in a loss of contributions from the rapidly decaying high-frequency components of the tip stray field distribution.

    \subsection*{Artificial QR Code Magnetic Pattern}
    
\begin{figure}
        \centering
        \includegraphics[width=0.49\textwidth]{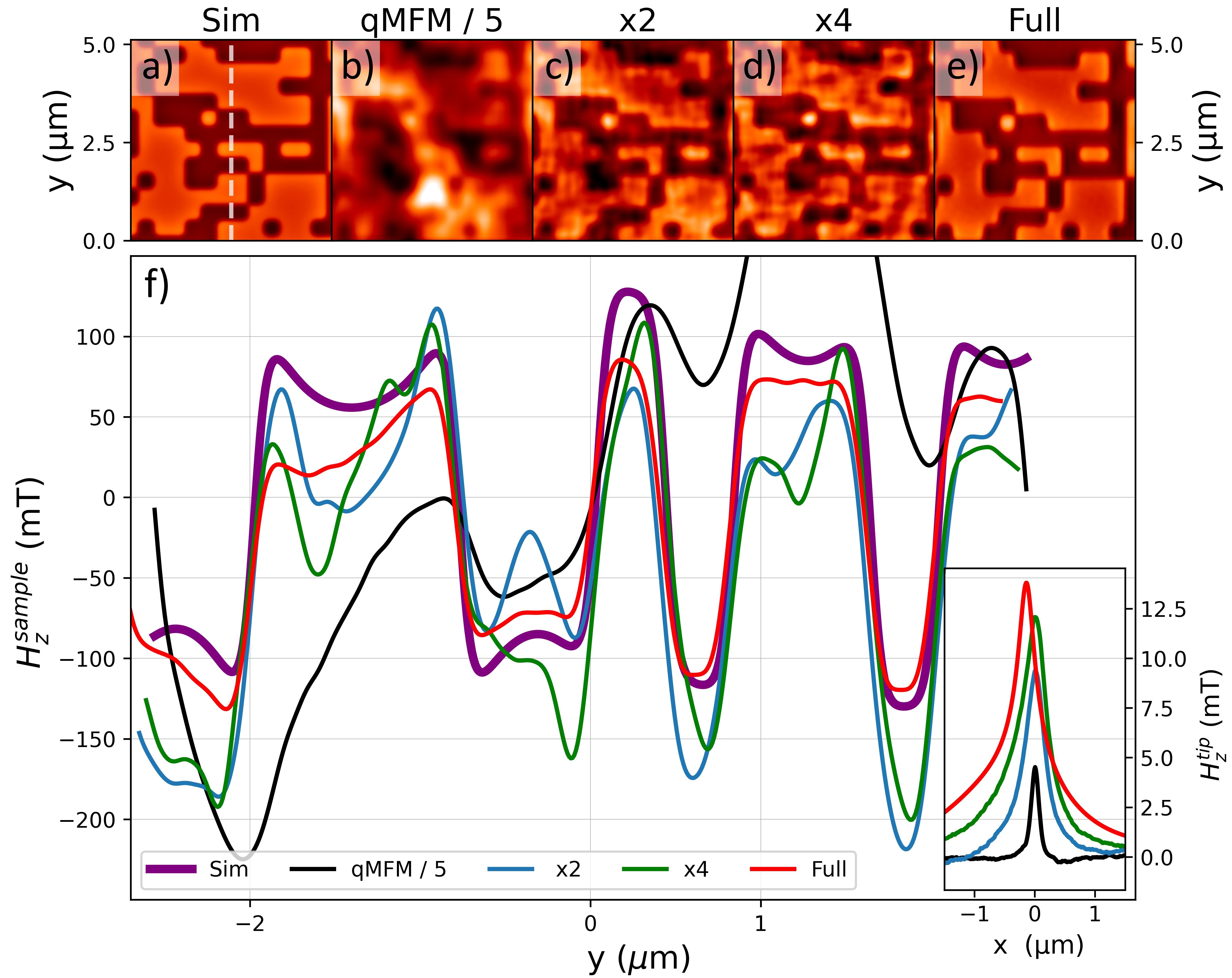}
        \caption{Artificial QR code sample: In the top row, the two-dimensional $\mu_0 H_z$ of the magnetic structure is shown, obtained by simulation (a), deconvolution using the transfer function from the original reference pattern (b) (note that this dataset is scaled by a factor of 0.2), the reference pattern with doubled domain size (x2, c), and the largest reference pattern (x4, d). The field in (e) is derived from deconvolution using the field distribution obtained from NV measurements. All plots, except for the qMFM data (factor 0.2), share the same color scale. In (f), plot lines \textcolor{reviewer1}{(along the dashed line)} representing the maximum values of the skyrmion field distributions are compared. The inset figure shows the transfer functions used for these deconvolutions.}
        \label{fig:validation_qr_structure}
    \end{figure}

    Figure~\ref{fig:validation_qr_structure} presents the derived stray field data $\mu_0 H_z$ for the QR code pattern. The worst result, which significantly deviates from the simulated stray field data, arises from deconvolution using the experimental TTF (Fig.~\ref{fig:validation_qr_structure}b and Fig.~\ref{fig:validation_qr_structure}f – black). Note, that in this figure, the derived amplitudes for this dataset had to be scaled by a factor of 0.2 (marked ``/5'' in Fig.~\ref{fig:validation_qr_structure}b) to plot them alongside the other results. In addition to this strong deviation of the field amplitude, the spatial structure of the simulated stray field is not well resolved and captures only a few basic features.   
    
    The $\mu_0 H_{z}^\text{tip}$ data reconstructed from TTFs derived from larger reference patterns ($\times 2$ and $\times 4$) are much more accurate (Figures~\ref{fig:validation_qr_structure}c, \ref{fig:validation_qr_structure}d, and \ref{fig:validation_qr_structure}f – blue and green) which can be attributed to the higher density of $k$-vectors in the relevant spatial frequency range. Note that the best reconstruction results from deconvolution using the $B_z^{\text{NV}}$ data  (Fig.~\ref{fig:validation_qr_structure}e and Fig.~\ref{fig:validation_qr_structure}f – red).

    \textcolor{reviewer1}{The significantly poorer performance of the QR code sample (compared to the artificial skyrmion sample) when deconvolving the sample via the experimental TTF can be explained by examining the Fourier spectra of both SUTs and the reference sample.}

    \subsection*{Fourier Spectra of Skyrmion-, QR-, and Reference-Sample}
    
\begin{figure}
        \centering
        \includegraphics[width=0.49\textwidth]{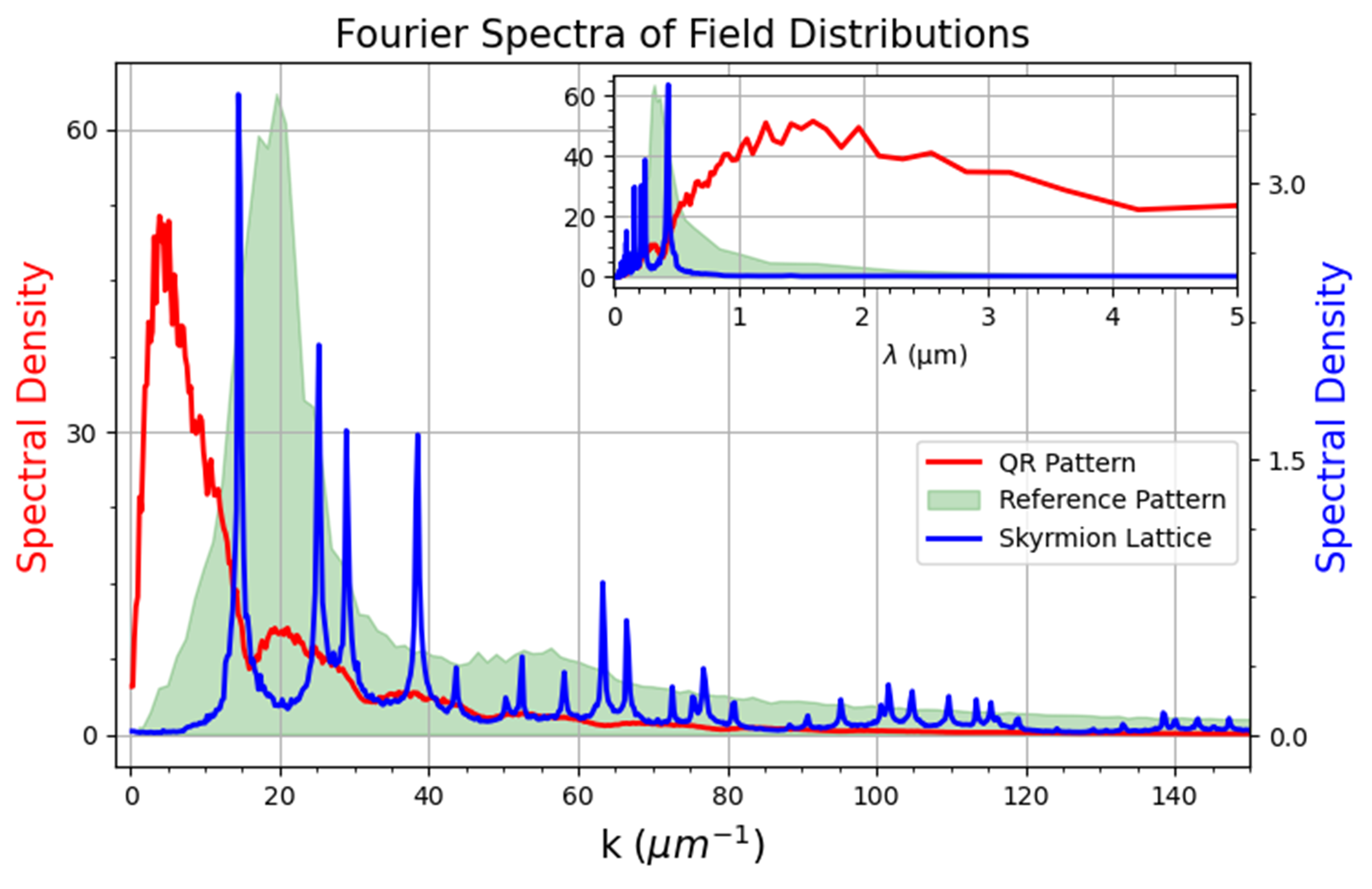}
        \caption{Comparison of the spectral density between test samples: a QR patterned magnetic structure (red) and a skyrmion lattice (blue), with the original reference pattern (green). The main figure is plotted in $k$-space, while the inset figure is presented in wave-vector space.}
        \label{fig:spectral_density_qr_skyrmion}
    \end{figure}

    Fig.~\ref{fig:spectral_density_qr_skyrmion} compares the spectra of the reference sample with our two test samples. Previously, the Fourier spectra were presented as a function of wavelength (shown in the inset of this figure); however, they are now plotted in the main figure as a function of wave vector to better resolve relevant features. The Fourier spectrum of the reference sample (green shading) shows good overlap with that of the skyrmion sample (blue) but lacks features at low wavelengths that are essential for accurately representing the QR code (red), which explains the observed results.
    
    To summarize, an ideal TTF should show contributions from a wide frequency spectrum to make it applicable to samples with a broad range of different feature sizes. This would, however, require reference samples that also contain a large spectrum of feature sizes, which are currently not available. Nevertheless, for accurately reconstructing an MFM-measured stray field with a limited feature size distribution, a reference sample with a limited spectrum is sufficient for the calibration, as long as it covers the dominant feature sizes of the sample under test. Accordingly, for the skyrmion sample, which features structure sizes around 100\,\text{nm}, the TTF calculated from experimental measurements of the CoPt reference pattern provided the best results. In contrast, the stray field of the QR structure, with its micron-sized features, was most accurately reconstructed using a TTF based on $B_z^{\text{NV}}$ data.
    
\section{\label{sec:summary}Summary and Conclusion}

    We calculated a set of tip transfer functions (TTFs) from simulated magnetic force microscopy (MFM) measurements of artificial reference samples with different characteristic feature sizes. These were compared to the TTF derived from a CoPt reference sample measurement and the tip stray field data obtained through quantum calibration using an NV center-based measurement. The simulations showed good agreement with experimental data, demonstrating the reliability of the approach. 
    
    However, both the simulated and measured TTFs based on the original pattern of the CoPt reference sample exhibit deficiencies in low-wavelength distributions, resulting in significant deviations from the $B_z^{\text{NV}}$ data. Moreover, while the NV-based measurement offers a quantum-based evaluation of the tip's stray field distribution, it lacks high-frequency contributions likely due to the interpolation model employed and the sparse density of measurement points near the peak of the stray field distribution. We assume that information about the stray field induced by magnetization near the tip apex is compromised. 
    
    TTFs are crucial for quantitative MFM measurements as they enable the calculation of the instrument calibration function (ICF), which is essential for deconvolving MFM data to determine the stray field distribution of a sample under test (SUT). Our findings indicate that an accurate reconstruction of a SUT's measured stray field distribution can be achieved if substantial overlap between the Fourier spectra of the SUT and the reference sample used for determining the TTF is ensured. Remarkably, even if a TTF significantly deviates from the actual tip stray field distribution, it can still yield very precise reconstructions, provided there is sufficient overlap. 
    
    Consequently, reference samples featuring maze domain patterns are particularly well-suited for analyzing magnetic structures with characteristic feature sizes on the 100\,\text{nm} scale. For larger feature sizes, new reference samples with micrometer-scale features need to be developed, tested, and validated. This will extend qMFM capabilities from the nanometer to the micrometer scale and help bridge the calibration gap to techniques with lower spatial resolution like magneto-optical indicator film measurements \cite{Dorosinskiy2023}.\\

\section*{Acknowledgments}

    We acknowledge Marcelo Jaime for critically reading the manuscript. We acknowledge funding from Deutsche Forschungsgemeinschaft (DFG, German Research Foundation) under Germany’s Excellence Strategy – EXC-2123 QuantumFrontiers – 390837967 and under the Priority Program Skyrmionics under grant SCHU 2250/8-1.

\section*{AUTHOR DECLARATIONS }

    \subsection*{Conflict of Interest}
        The authors have no conflicts to disclose.

    \subsection*{Author Contributions}

    \textbf{Baha Sakar:} Conceptualization (equal); Data curation (lead); Formal analysis (equal); Investigation (equal); Methodology (equal); Project administration (support); Resources (support); Software (lead); Supervision (equal); Validation (lead); Visualization (equal); Writing - original draft (equal); Writing – review \& editing (equal). 
    
    \textbf{Christopher Habenschaden:} Conceptualization (support); Data curation (equal); Formal analysis (equal); Investigation (equal); Methodology (equal); Project administration (support); Resources (support); Software (support); Supervision (support); Validation (equal); Visualization (equal); Writing - original draft (support); Writing – review \& editing (equal). 
    
    \textbf{Sibylle Sievers:} Conceptualization (equal); Data curation (supporting); Formal analysis (equal); Funding acquisition (equal); Investigation (equal); Methodology (equal); Project administration (lead); Resources (equal); Software (equal); Supervision (equal); Validation (equal); Visualization (support); Writing - original draft (equal); Writing – review \& editing (equal).
    
    \textbf{Hans Werner Schumacher:} Conceptualization (equal); Funding acquisition (equal); Project administration (lead); Resources (equal); Supervision (equal); Writing - original draft (support); Writing – review \& editing (support).

\section*{Data availability statement}

    The data that support the findings of this study are available from the corresponding author upon reasonable request.

\section*{References}

\nocite{*}

\bibliography{main}

\end{document}